\documentclass[journal=aelccp,manuscript=letter]{achemso}

\usepackage[version=3]{mhchem} 
\usepackage[T1]{fontenc} 
\usepackage{multirow}
\usepackage{amsmath, amssymb}
\usepackage{graphicx}
\usepackage{caption} 
\usepackage{makecell}

\usepackage{color}

\author{Luo Yan}
\affiliation{Institute of Fundamental and Frontier Sciences, University of Electronic Science and Technology of China, Chengdu 610054, P. R. China. }
\author{ Tao Bo}
\affiliation{Songshan Lake Materials Laboratory, Dongguan, Guangdong, 523808, China.}
\author{ Bao-Tian Wang}
\affiliation{Spallation Neutron Source Science Center, Institute of High Energy Physics, Chinese Academy of Sciences, Dongguan 523803, China.}
\alsoaffiliation{Collaborative Innovation Center of Extreme Optics, Shanxi University, Taiyuan, Shanxi 030006, China.}
\author{Sergei Tretiak}
\affiliation
{Theoretical Physics and Chemistry of Materials, Los Alamos National Laboratory, Los Alamos, New Mexico 87545, USA.}
\alsoaffiliation{Skolkovo Institute of Science and Technology, Moscow 143026, Russia.}
\author{Liujiang Zhou}
\affiliation{Institute of Fundamental and Frontier Sciences, University of Electronic Science and Technology of China, Chengdu 610054, P. R. China. }
\email{liujiang86@gmail.com}

\title[An \textsf{achemso} demo]
  {Single-Layer Di-titanium oxide Ti$_{2}$O MOene:  Multifunctional Promises for Electride, Anode Materials, and Superconductor}

\abbreviations{IR,NMR,UV}
\keywords{MOene, Anode  material, Electride, Superconductivity, First-principles calculations}

\begin{document}

\begin{abstract}
 Using the first-principles calculations, we report the existence of  the single-layer (SL) di-titanium oxide  Ti$_2$O (labeled as MOene) that constructs a novel family of MXene based on transition metal  oxides. This MOene material strongly contrasts the conventional ones consisting of transition metal  carbides and/or nitrides. SL Ti$_2$O has high thermal  and dynamical stabilities  due to the strong Ti$-$O ionic bonding interactions. Moreover, this material is an intrinsic electride and exhibits extremely low diffusion barriers of $\sim$12.0 and 6.3  meV for Li- and Na diffusion, respectively. When applied as anode materials in  lithium-ion batteries and sodium-ion batteries, it possesses a high energy storage capacity (960.23 mAhg$^{-1}$), surpassing the traditional MXenes-based anodes. The superb electrochemical performance stems from the existed anionic electron on Ti$_2$O surface. Astonishingly, SL Ti$_{2}$O is also determined to be a superconductor with a superconducting transition temperature (\textit{T$_{c}$}) of $\sim$9.8 K, which originates from the soft-mode of the first acoustic phonon branch and enhanced  electron-phonon coupling   in the low-frequency region. Our finding broadens the family of MXenes and would facilitate more experimental efforts toward future nanodevices.
\end{abstract}

\section{Introduction}
Recently, two-dimensional (2D) metal-shrouded  crystals (MSC) (M$_2$X, M = metal and X = C, N) have  triggered enormous interest due to their excellent mechanical and electronic (metallic and/or semiconducting) properties, providing promises for sensors,\cite{liu2015novel,yu2015monolayer} catalysts,\cite{gao2014preparation,seh2016two,xiong2018recent} energy storage,\cite{anasori20172d,sun2018two} (topological) electrides,\cite{huang_topological_2018,zhou_interlayer-decoupled_2018} superconductors,\cite{xu_large-area_2015,lei2017predicting,zhang2017superconductivity} \textit{etc}. These MSC have a trigonal structure with two layers of M atoms covering one layer of X atoms.  This family encompasses both non-transition and transition metal shrouded carbides or nitrides. The former includes alkali-metal and alkali-earth-metal carbides and nitrides (e.g., Ca$_{2}$N,\cite{zhao2014obtaining} Mg$_{2}$C\cite{wang_monolayer_2018}), which were mainly investigated by theory.  The latter  refers to transition-metal shrouded  MXenes (TMSM), including Ti$_{2}$C,\cite{naguib2012two} Mo$_{2}$C,\cite{meshkian2015synthesis} Ti$_{2}$N, \cite{soundiraraju2017two} V$_{2}$C, \cite{liu2017preparation} Nb$_{2}$C,\cite{naguib2013new}  Y$_2$C,\cite{huang_topological_2018} etc., which have been widely explored over the past few years. 

Following remarkable promises of conventional MSM, metal-shrouded  monochalcogenides, such as single-layer (SL) Tl$_{2}$O,\cite{ma_single-layer_2017}  A$_2$X (A = Na, K, Rb, or Cs; X = O, S, Se, or Te),\cite{hua_dialkali-metal_2018}  KTlO,\cite{song2019ktlo:} and In$_{2}$X (X = O, S, Se),\cite{li_first-principles_2020}  have also showed fascinating physical and chemical traits, substantially enriching the family of 2D MSC.  Considering  the potential  uses, the toxicity and scarcity of element Tl (as rare as gold) and the highly activity of alkali metals when exposed to  ambient conditions, seem to be detrimental. In contrast, Ti is a transition metal  with a high abundance in the earth's crust\cite{tan_abundance_1970} and multifarious uses in the modern day technologies. 
It is worthy noted that the bulk Ti$_{2}$O phase in a hexagonal lattice has been accessed in experiments  since 1970,\cite{kornilov_neutron_1970, Novoselova2004} motivating us to explore the existence of  SL  or few-layer Ti$_{2}$O with atomic thickness. Addressing this issue would not only  open the door to a new family of MXenes, i.e.,  metal-shrouded 2D transition metal oxides, but also potentially enrich the physics of 2D materials.

In this work, the  first-principles calculations  suggest that  SL di-titanium oxide Ti$_{2}$O is a novel family of MXene based on transition metal oxides (labels as MOene to underline the 2D morphology)  rather than conventional ones consisting of transition metal carbides and/or nitrides. This MOene is a stable material and an intrinsic electride  with anionic electron on the surface. The anionic states  of  2D Ti$_{2}$O  enable the very low diffusion barrier and large charge storage capacity when explored as an anode material of lithium-ion batteries (LIBs) and sodium-ion batteries (SIBs). Meanwhile, this  MOene is a 2D superconductor with a  \textit{T$_{c}$} of $\sim$9.8 K.  Such a combination of stability and exceptional electronic features implies a number of potential applications of Ti$_{2}$O in the next-generation nanodevices.

\section{Computational Methods}
\par First-principles calculations based on the density functional theory (DFT) calculations were performed  the projector augmented wave (PAW) scheme. The exchange and correlation contributions were simulated within the generalized gradient approximation (GGA) \cite{blochl1994projector,blochl1994improved} as formulated by Perdew-Burke-Ernzerhof (PBE).\cite{perdew1996generalized} A rapidly dispersion-corrected DFT method (opt88-vdw) was adopted to model the van der Waals interaction.\cite{klimes2011van}   A subset of numerically expensive calculations were performed with an accurate screened exchange hybrid density functional by HSE06\cite{heyd2004efficient} and inclusion of spin-orbit coupling (SOC) terms. The plane-wave cutoff energy of 550 eV and $\Gamma$-centered 21 $\times$ 21 $\times$ 1 \textit{k}-point mesh using the Monkhorst-Pack method were adopted in electronic structure calculations. To avoid the interactions between the nearest-neighbor unit cells, a vacuum thickness of 20 \AA{} along the \emph{z} direction was applied. The geometry optimizations were carried out with a convergence threshold of  10$^{-5}$ eV in total energy and of 0.01 eV {\AA}$^{-1}$ in maximum force per each atom. Some data analysis were done with the help of VASPKIT.\cite{wang2020vaspkit} See more details in Supporting Information. 

\section{Results and discussion}

Figure 1a shows the crystal model of SL Ti$_{2}$O. The SL structure optimizes in a hexagonal lattice with a space group \emph{P$\bar{3}$m1} (no. 164) and has a  D$_{3d}$ point symmetry. This structure is typical for SL MSM featuring the central oxygen octahedrally coordinated to six Ti atoms. Obviously, it breaks the traditional definition for MXenes, where X = C, N.  Also, it is different from SL MBene, where X = B. \cite{ozdemir2019a} Here, we take SL Ti$_{2}$O as an example to propose a novel family of 2D material, MOene, where O atomic layer is sandwiched by two transition metal layers. To clearly determine its ground state, four types of spin oriented configurations are considered, \textit{i.e.}, ferromagnetic (FM), antiferromagnetic Néel (AFM-Néel), antiferromagnetic Stripy (AFM-Stripy) and antiferromagnetic zigzag (AFM-Zigzag)  (Figure 1b).  Their corresponding total energies and magnetic moments of each Ti atom are listed in Table I. Intriguingly, FM, AFM-Stripy, and AFM-Zigzag tend to be nonmagetic system with the same energy and zero magnetic moment on per Ti atoms. However, the  AFM-Néel is the most stable configuration and has a magnetic moment 0.76 $\mu$$_{B}$ per Ti atom, arising from the dangling 3\textit{d} orbitals of  Ti atoms. Thus, the AFM-Néel configuration is used in the next simulations. The equilibrium  lattice parameters are \textit{a} = 2.82 {\AA}, \textit{b} = 4.88 {\AA}, which is in accordance with experimental values of Ti$_{2}$O film (\textit{a} = 2.84 {\AA}).\cite{fan2019structure} The  electron localization function (ELF) plot reveals the vanishing electron distributions between Ti and O atoms (Figure 2a), suggestive of the ionic bonding interactions via which the charge transfer are from Ti to more electronegative O atoms (Figure S2a$-$b).

\begin{table}[htbp]
	\small
	\caption{\ The total energy (eV) of per unit cell and magnetic moment ($\mu$B) of per Ti atom for different configurations of SL Ti$_{2}$O. }
	\label{tbl:tbl-2}
	\begin{center}
		\begin{tabular}{cccccccccccccccc}
			\hline
			\hline
			Configuration & Energy (eV) & Magnetic moment ($\mu$B)&\\
			\hline
			Nonmagnetic & -35.59192 & 0 &\\
			FM & -35.59188 & 0 &\\
			AFM-Néel &-35.83279& 0.76& \\
			AFM-Stripy & -35.59192 & 0 &\\
			AFM-Zigzag &-35.59191& 0 &\\
			\hline
		\end{tabular}
	\end{center}
\end{table}

The dynamical stability of SL Ti$_{2}$O with AFM-Néel configuration is confirmed  by the  phonon spectrum lacking unstable modes (Figure S1a). The \textit{ab initio} molecular dynamics (AIMD) simulations were carried out next to assess the thermal stability.  SL Ti$_{2}$O is stable without Ti$-$O bonds breakage and geometric reconstructions  up to 700 K (Figures S1b,c). The elastic constants are calculated  to be C$_{11}$ = C$_{22}$ = 133.7 N/m, C$_{12}$ = 33.7 N/m,  C$_{66}$ = 6.2 N/m, respectively, satisfying the mechanical stability criteria for rectangle unit cell.\cite{born1954dynamical,mouhat2014necessary} The in-plane Young's modulus (Y$_{x}$ = Y$_{y}$) are 125.2 N/m. The Poisson's ratios along \textit{x} and \textit{y} directions ($\nu$$_{x}$ and $\nu$$_{x}$) are both determined to be 0.25 due to the isotropic nature. The cohesive \emph{E}$_{coh}$  and formation \emph{E}$_{f}$ energies are 5.20 and $-$0.44 eV/atom, respectively, indicating its exothermic feature in the chemical  synthesis and high experimental feasibility. Fortunately, the SL Ti$_{2}$O may be accessed via the chemical deposition on (0001)-oriented $\alpha$-Al$_{2}$O$_{3}$ single crystalline substrate,  as observed in  Ti$_{2}$O thin films.\cite{fan2019structure}

Ti$_{2}$O monolayer is expected to accommodate the anionic  states, as observed in other SL TMSM.\cite{hu20152d,Lee2013,park2017strong,Zhou2018} As shown in ELF plot (Figure 2a), there are two excess electron pools localized above the SL Ti$_{2}$O surfaces as verified by the de-localized electride with anionic electrons. The chemical formula can thus be expressed as [Ti$_{2}$O]$^{2+}$$\cdot$2e$^{-}$, similar to the Y$_{2}$C electride.\cite{park2017strong} The difference charge plot show the charges  are mainly transferred  from less electronegative Ti   to more electronegative O atoms and the center sites of anionic electrons (denoted as  X  pseudo atoms ) that  reside above the center of Ti atoms (Figure S2). The anionic electron bands (X-bands)  are mostly located  below the Fermi level (Figure 2b), in accordance with  the  partial density of states (PDOS) in the  range of $-$3 eV $<$ E$-$E$_{F}$ $<$ 0 eV (Figure 2c). Meanwhile, SL Ti$_{2}$O has a  metallic feature with substantial electron states crossing the Fermi level. As indicated in Figure 2c,  the dominant contributions to the electronic states near the Fermi level  mainly stem from  the hybridization between Ti-\textit{d}$_{x^{2}-y^{2},xy}$ and Ti-\textit{d}$_{xz,yz}$ orbitals.  The anionic electron sites have  obvious hybridization with Ti-3\textit{d} orbitals in the  range of $-$3 eV $<$ E$-$E$_{F}$ $<$ 0 eV (Figure 2b and Figures S3a,b). Therefore, anionic electrons are significantly from the unsaturated Ti-3\textit{d} orbitals. To clearly determine the electronic states in Ti$_{2}$O system, we further checked it in bulk Ti$_{2}$O. Intriguingly,  the electronic states intrinsically reside within the interlayers (Figure S4), which helps to protect the electride states  from  oxygen or water contamination that destroys the anionic electride states  of SL Ti$_{2}$O.

Since the electronic structures of conventional MXene are usually affected by surface functional groups, we here probe the   functionalization on SL Ti$_{2}$O via fluoridation, hydrogenation, hydroxylation, and oxidization. Here, the  four major possible functionalization configurations  are considered (Figures S5,6). After full structure optimizations, the F and OH groups  locating on the top of O atoms are most energetically  favorable (Figure S5c), while the  most stable configurations for H and O groups  locate above the hollow site of hexatomic ring of Ti and O atoms (Figure S6b). All of the functionalized Ti$_{2}$O show nonmagnetic ground-state due to the fact that the dangling 3\textit{d} orbitals of the Ti atoms are saturated by corresponding  functional groups. The phonon spectraconfirm that the fluorinated, hydrogenated, hydroxylated SL Ti$_{2}$O are dynamically stable, while oxidized Ti$_{2}$O contains the unstable imaginary modes (Figure S7). The optimized lattice constants for Ti$_{2}$OF$_{2}$, Ti$_{2}$OH$_{2}$ and Ti$_{2}$O(OH)$_{2}$ are $\textit{a}$ = $\textit{b}$ = 2.85, 2.83 and 2.89 {\AA}, respectively. The band structures of  Ti$_{2}$OF$_{2}$, Ti$_{2}$OH$_{2}$ and Ti$_{2}$O(OH)$_{2}$ are presented in  Figures 2d and S8. Ti$_{2}$OH$_{2}$ and Ti$_{2}$O(OH)$_{2}$ are semi-metal with finite electronic states crossing the Fermi level. In contrast, Ti$_{2}$OF$_{2}$ exhibits  a semiconductor trait with  a direct energy gap (E$_{g}$) of $\sim$0.5 (PBE) and 1.1 eV (HSE) (Figure 2d), displaying great potentiality in optoelectronic devices.  Thus, the electronic properties of SL Ti$_{2}$O  can be efficiently  tuned upon varying the surface functional groups.

Considering the promising application of conventional MXene materials  in  LIBs and SIBs,\cite{tang2012mxenes,hu2014investigations,sun2016ab,yu2018tic3,wang2017first} we here explore the feasibility of  Ti$_{2}$O MOene as a suitable electrode for the rechargeable batteries. The  capability of energy storage for Ti$_{2}$O monolayer was investigated by the adsorption energies. Three typical Li/Na adsorption sites (labeled as S1-S3) were considered (Figure 3a).  Calculated adsorption energies  of Li/Na adsorption identified S3 site to be the most stable site with the lowest adsorption energies of being $-$0.75 and $-$0.96 eV for Li and Na atoms, respectively (Figure 2b). These adsorption energies are lower than those on Mo$_{2}$C ($-$0.58 eV for Li, $-$0.77 eV for Na) \cite{sun2016ab}, Ti$_{3}$C$_{2}$ ($-$0.50 eV for Li), \cite{tang2012mxenes} TiC$_{3}$ ($-$0.50 eV for Na)\cite{yu2018tic3} and V$_{2}$C ($-$0.16 eV for Li), \cite{hu2014investigations} and comparable to Ti$_{2}$N ($-$0.75 eV for Li, $-$0.95 eV for Na) monolayers,\cite{wang2017first}  indicating the strong interactions between the Li/Na atom and Ti$_{2}$O. Based on Bader analysis,\cite{tang2009a} the net charge transfer from the adsorbed Li and Na atoms on the S3 site to the Ti$_{2}$O monolayer are 0.87 and 0.73 e/atom (Figures S9a,b), respectively, suggesting the ionic bonding feature stemming from the \textit{s}$-$\textit{d} hybridization between the Ti-3\textit{d}  and the metal \textit{s} orbitals. Additionally, the Li and Na adsorbed systems maintain the metal trait (Figures S9c,d), indicating the good electronic conductivity in SL Ti$_{2}$O.

The charge-discharge rate was  assessed via the energy barrier of Li/Na ion diffusion on Ti$_{2}$O monolayer using the  climbing-image nudged elastic band (CI-NEB) method\cite{mills1994quantum}. As shown in Figure 3c, the  energy barrier  for Li (Na) ion is low to be about 12 (6.3) meV along the path I or III (I). The extremely low energy barrier indicates an ultra-fast  fast charge/discharge process when SL Ti$_{2}$O used as the anode materials. Such low diffusion barriers for Li and Na ions are lower than the previous  values in other MXenes ($\sim$14$-$70 meV\cite{sun2016ab,tang2012mxenes,wang2017first}), suggesting an ultra-efficient Li/Na ion diffusion and thus ultra-fast  fast charge/discharge process  on the SL Ti$_{2}$O when used in rechargeable batteries. This unprecedentedly small diffusion barrier can be attributed to the surface-confined anionic electrons, which helps to smooth the surface potential effectively, as confirmed in  SL Ca$_{2}$N electride.\cite{hu20152d}

Four  configurations of Li/Na adducts (\textit{i.e.}, Ti$_{2}$OM$_{n}$(M = Li, Na), n = 1$-$4) (Figures S10a-d) on a 2 $\times$ 2 $\times$ 1 super-cell were considered to evaluate the  storage capacity via the concentration-dependent adsorption behaviors.   All considered mixtures are  energetically stable, confirmed by the negative formation energy and  convex hulls (Figures 4a,c).
Additionally, to estimate the  volume effect, the geometries after full adsorption of Li and Na atoms are checked. The thicknesses of   SL Ti$_{2}$O host after full adsorption of Li and Na atoms are 2.41 and 1.87 {\AA} for the two-layer Li- and Na-adsorbed configurations, respectively. Compared with the pristine monolayer (2.82 {\AA}), the volume of the fully Li (Na) adsorbed systems increases slightly to 2.73 (7.17) \%. Meanwhile, the bond length of Ti$-$O  has only a small change of $\sim$1.4 \% for adsorbed mixtures. The layered E$_{ave}$ are calculated to be $-$0.57  and $-$0.06 eV per atom for the one and two-layer metal adsorption. Such a low absorption energy for the two-layer adatoms indicates a weak   adsorption interactions. However, this value is still smaller than those of Mo$_{2}$C ($-$0.01 eV per atom)\cite{sun2016ab} and Nb$_{2}$C($-$0.02 eV per atom), \cite{hu2016investigations} and comparable to that V$_{2}$C ($-$0.06 eV per atom) \cite{hu2014investigations} in LIBs. The layered E$_{ave}$ for the one- and two-layer Na adatoms are   $-$0.27 and $-$0.03 eV per atom. It should be worthy noted that the two-layer E$_{ave}$ is still smaller than that of typical electrode materials in SIBs, such as MoN$_{2}$ ($-$0.02 eV per atom),\cite{zhang2016theoretical} Ca$_{2}$N (-0.003 eV per atom)\cite{hu20152d} and GeS ($-$0.02 eV per atom). \cite{li2016germanium}  Therefore, the Li and Na atoms in the corresponding layer prefer to chemically bind to the host material (negative E$_{ave}$), rather than to form metal clusters  (positive E$_{ave}$).\cite{sun2016ab,hu2016investigations,zhang2016theoretical} The theoretical capacities $C_A$  for Ti$_{2}$O in LIBs and SIBs are  960.23  mAhg$^{-1}$,  larger than those of conventional MXenes (100$-$542 mAhg$^{-1}$).\cite{xie2014role,sun2016ab,wang2017first,hu2016investigations} The ELF plots (Figure S11) upon Li and Na atoms adsorption  reveal that the main origin of high capacity behavior is from the anionic electrons, which stabilize adsorbed Li/Na ions and favor the high adsorption concentration of Li/Na atoms.\cite{xie2014prediction,xu2019structural}

In addition, the average open-circuit voltage (OCV) in the range of 0$-$1.0 V indicates the ability to   prevent the  dendrite formation of alkali metals during the discharge/charge process. \cite{yu2018tic3,eames2014ion,wang2017first} There are four main plateaus during the Li (Na)  insertion process (Figures 4b,d), showing a steady decreasing trend from 0.64 to 0.28 V (0.63 to 0.13 V) upon increasing the concentration of  Li(Na) adatoms. Moreover, the OCV for full sodiations of Ti$_{2}$OLi$_{4}$ and Ti$_{2}$ONa$_{4}$ are 0.28 and 0.13 V, respectively, which are rather low and comparable to the typical anode materials in LIBs and SIBs, such as Mo$_{2}$C,\cite{sun2016ab} MoS$_{2}$,\cite{hu2014mos2} TiC$_{3}$ \cite{yu2018tic3} and Ti$_{3}$C$_{2}$ \cite{er2014ti3c2} \textit{etc}. Therefore, given the above discussions, the SL Ti$_{2}$O should be a excellent candidate as anode material for rechargeable LIBs and SIBs.

The superconductivity of the metallic SL Ti$_{2}$O was further investigated. Figure 5a shows the resolved  phonon spectra in terms of the displacement directions of Ti and O atoms,  in agreement with the projected phonon density of states (PhDOS) (Figure 5b). The  out-of-plane modes of Ti atoms (Ti-\emph{z}) dominate the low-frequency region (below 200 cm$^{-1}$). The mid-frequency region from 200 to 300 cm$^{-1}$ mainly consists of in-plane vibrations of Ti atoms.  Figure 5c  shows that the   relatively large  strength of the electron-phonon coupling (EPC)   (labeled by $\lambda$$_{\textbf{q}v}$  ) are related to the soft-modes in the first acoustic branch with a frequency of $\sim$150 cm$^{-1}$. This soft-mode that stems from the Ti-\emph{z} vibrations makes a significant contribution to the cumulative frequency-dependent  EPC function $\lambda$($\omega$). The Eliashberg spectral function $\alpha$$^{2}$F reveals that the two major peaks   in the low frequency region of 100$-$200 cm$^{-1}$ lead  to a rapid increase of the cumulative $\lambda$($\omega$) (Figure 5d), evidencing the jump of the total  EPC in this range and pointing  to a medium-coupling superconductor with a $\lambda$ = 0.79.   Based on the McMillian-Allen-Dynes formula,\cite{allen1975transition} the superconducting transition temperature \textit{T$_{c}$} is evaluated to be $\sim$9.8 K,  higher than that of  Mo$_{2}$C (3.0 K), \cite{xu_large-area_2015,lei2017predicting} Ti$_{2}$C (1.3 K)\cite{lei2017predicting} and functioned Nb$_{2}$C (4.5$-$7.1 K).\cite{Kamysbayev2020} This superconductivity in  Ti$_{2}$O MOene enriches the 2D MXene-based superconductors that are only observed in  Mo$_{2}$C, Ti$_{2}$C and Nb$_{2}$C materials.\cite{xu_large-area_2015,lei2017predicting,Kamysbayev2020}

\section{Conclusions}
\par In summary, we report a systematically theoretical investigation of a new SL metal-shrouded MOene,  namely, 2D di-titanium oxide Ti$_{2}$O, in great contrast conventional MXenes that consists of transition metal  carbides and/or nitrides.  This thermally and dynamically stable MOene is an  electride in monolayer and bulk form. Ti$_{2}$O MOene can be used as an anode material in lithium and sodium battery technologies, owing to extremely low diffusion barriers (12.0  meV for LIBs and 6.3  meV for NIBs) and an excellent storage capacity (960.23 mAhg$^{-1}$). The   anionic electrons intercalating the adsorbed Li/Na atoms and Ti$_{2}$O monolayer make the significant contributions to the very fast charge/discharge process. Moreover, within the Bardeen-Cooper-Schrieffer microscopic theory, \cite{bardeen1957theory,Zhang2017,grimvall1981electron,giustino2017electron} Ti$_{2}$O is determined to be a intrinsic superconductor with a transition temperature \emph{T$_{c}$} of 9.8 K. The superconductivity mainly originates from the presence of soft modes in the first acoustic branch and a significant enhancement of EPC in the low-frequency region. These  results endow  Ti$_{2}$O MOene with   multi-functionality, promisingly suitable for future nano-devices and energy storage technologies, and thus call for further experimental fabrication and characterization efforts.

\section*{Conflicts of interest}
There are no conflicts to declare.

\section*{Acknowledgements}
L.Z. acknowledges the financial support from the University of Electronic Science and Technology of China. This work was performed in part at the Center for Integrated Nanotechnology (CINT), a U.S. Department of Energy and Office of Basic Energy Sciences user facility. B.-T.W. acknowledge financial support from the Natural Science Foundation of China (Grants No. 11675195 and No. 12074381. S. T. acknowledges support from the Los Alamos National Laboratory (LANL) Directed Research and Development funds (LDRD).


%
%

\bibliography{MOene}

\clearpage

\begin{figure}[hb!]
	\centering
	\includegraphics[width=1\linewidth]{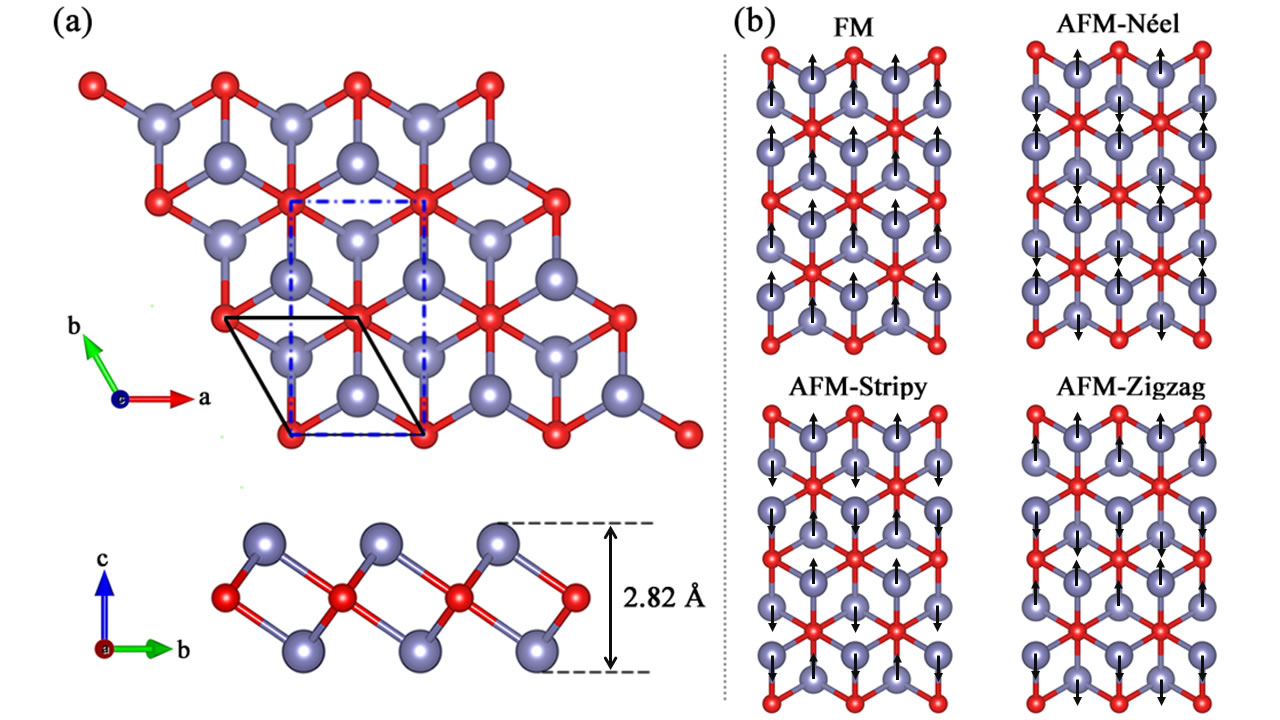}
	\caption{(a) Top (upper panel) and side (lower panel) views for SL Ti$_{2}$O. The rhombic primitive and conventional cells are labeled by  black sold and blue dash lines, respectively. Silvery gray: Ti; red: O.  (b) Four magnetic  configurations of SL Ti$_{2}$O. The up and down arrows indicate spin-up and spin-down orientations, respectively.}
\end{figure}

\clearpage
\begin{figure*}[hb!]
	\begin{center}
		\includegraphics[width=\linewidth]{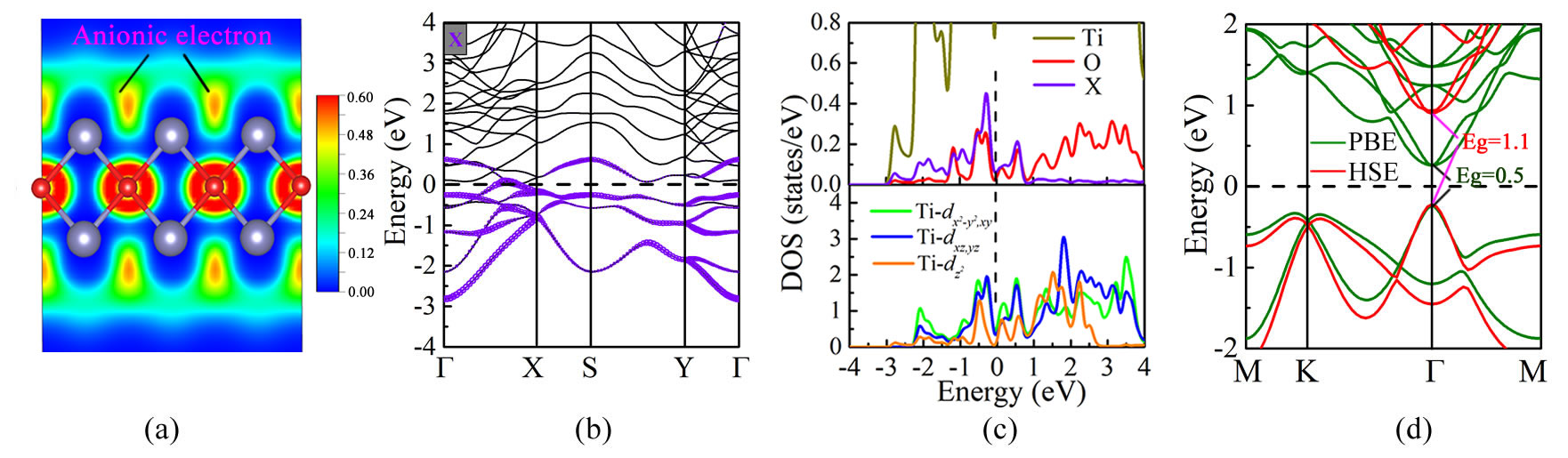}
	\end{center}
	\caption{(a) The ELF slice of Ti$_{2}$O monolayer from (0$\bar{1}$$\bar{2}$) plane. The isovalue used is 0.4 a.u. (b) The projected spin-up band structure  with the  anionic electrons highlighted in purple. (c) The spin-up PDOS of the Ti$_{2}$O monolayer.  (d) The band structure of fluorinated SL Ti$_{2}$O within PBE and HSE. The Fermi levels are set to be zero.}
	\label{charge}
\end{figure*}
\clearpage
\begin{figure*}[hb!]
	\begin{center}
		\includegraphics[width=\linewidth]{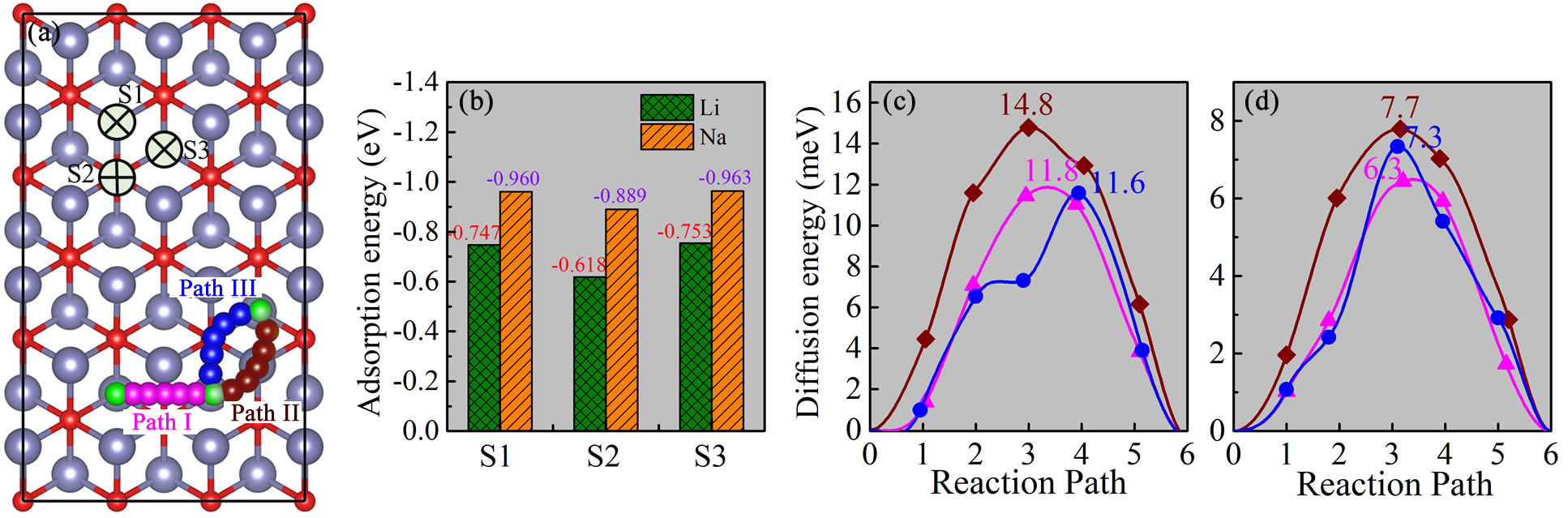}
	\end{center}
	\caption{(a) The three Li/Na  adsorption sites on  Ti$_{2}$O monolayer: S1, S2 and S3. S1 and S2 are located on the top of Ti and O atoms, respectively; S3 is the hollow site of hexatomic ring consisting of Ti and O atoms. Three possible migration pathways I, II, III near the neighboring low-energy adsorption sites are marked. (b) The calculated adsorption energies of different sites on SL Ti$_{2}$O. The corresponding computed diffusion barriers of the (c) Li and (d) Na migration along path I, II and III  on Ti$_{2}$O monolayer, respectively.}
\end{figure*}
\clearpage
\begin{figure}[hb!]
	\centering
	\includegraphics[width=0.8\linewidth]{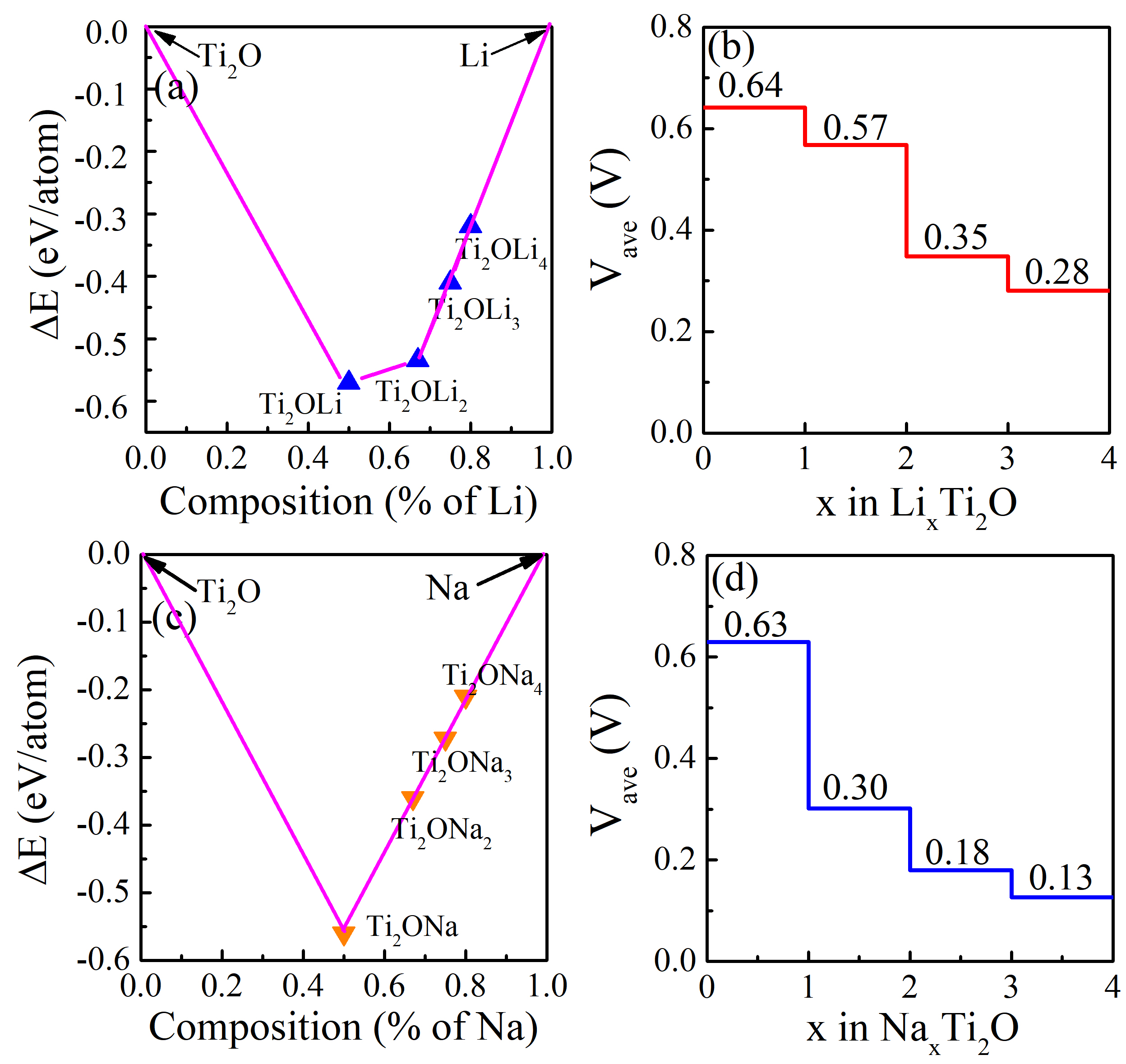}
	\caption{ The hull of formation energies for the most stable configurations of (a) Li- and (c) Na-adsorbed Ti$_{2}$O monolayer. The OCV for each (b) Li and (d) Na concentration, respectively. }
\end{figure}
\clearpage
\begin{figure}[hb!]
	\centering
	\includegraphics[width=0.5\linewidth]{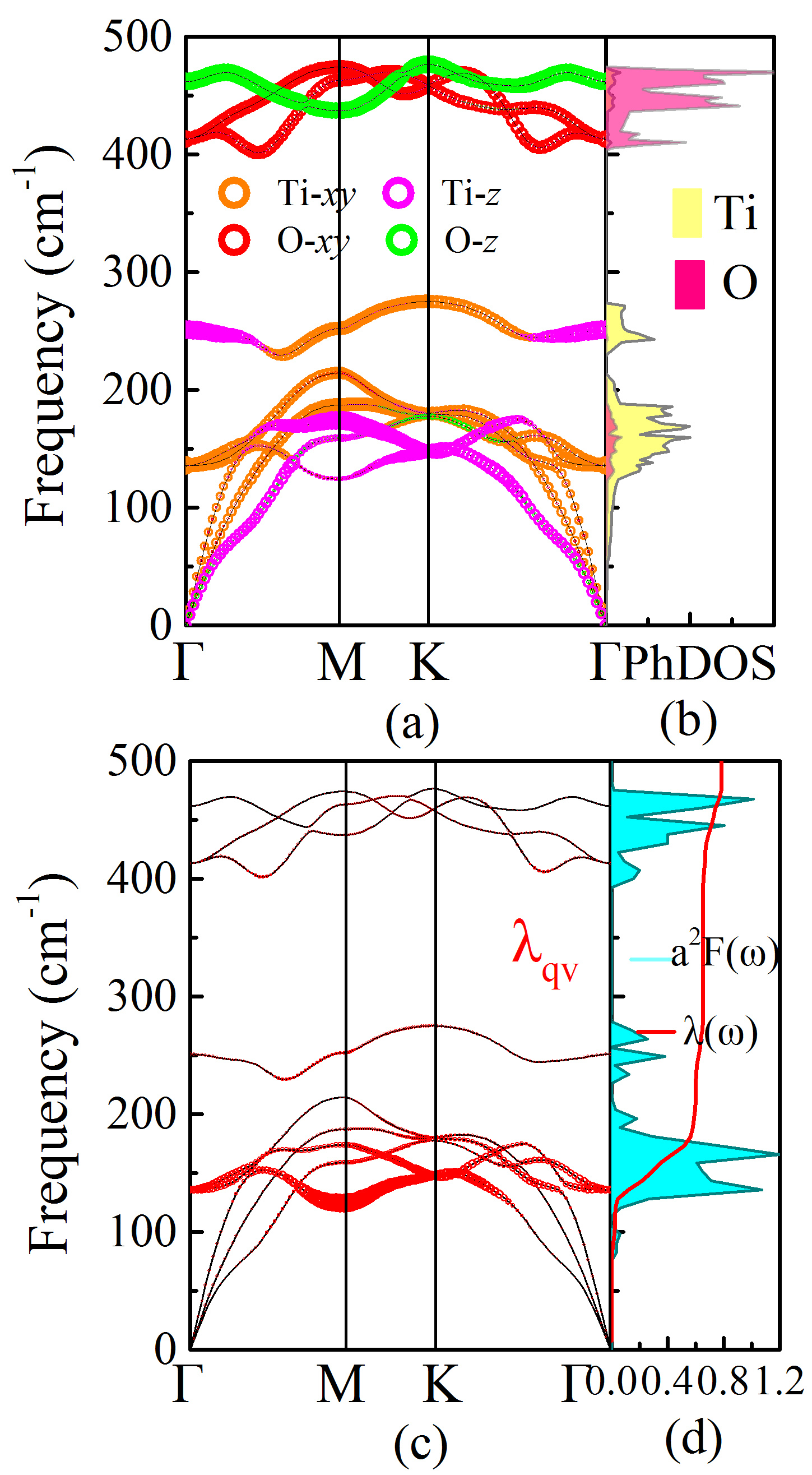}
	\caption{(a) The phonon spectra of Ti$_{2}$O resolved in terms of the vibration directions of Ti and O atoms. The orange, pink, red and green hollow circles indicate Ti in-plane, Ti out-of-plane, O in-plane and O out-of-plane modes, respectively. (b) PhDOS for Ti$_{2}$O. (c) The magnitude of the EPC $\lambda_{\textbf{\emph{q}}\nu}$.  (d)The Eliashberg spectral function $\alpha^{2}$F($\omega$) and the cumulative frequency-dependent EPC $\lambda$($\omega$).}
\end{figure}

\clearpage

\begin{tocentry}

	\centering
	\includegraphics[width=1\linewidth]{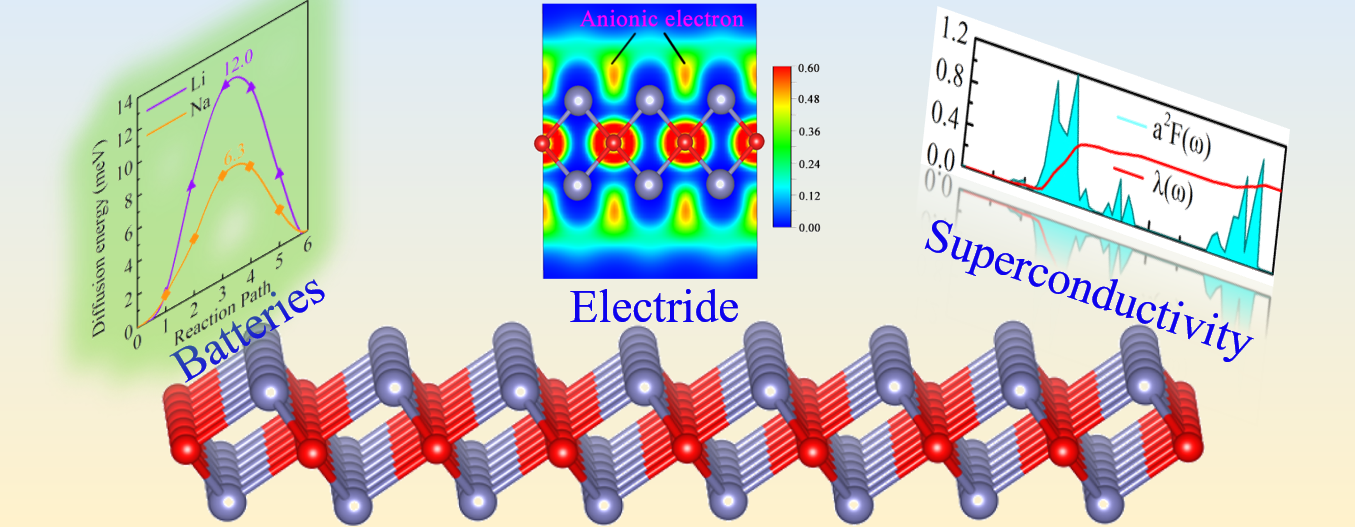}
\end{tocentry}

\end{document}